\documentclass[12pt,a4paper]{article}
\usepackage{amsmath}
\usepackage{stmaryrd}
\usepackage[dvipsnames]{xcolor}

\usepackage[left=3cm, right=3cm, top=3cm, bottom=3cm]{geometry}

\usepackage{eurosym}
\usepackage{lscape}

\usepackage{booktabs}
\usepackage{tabularx}
\usepackage{makecell}
\usepackage[flushleft]{threeparttable}
\usepackage{float}
\usepackage{caption}
\usepackage{longtable}
\usepackage{multirow}

\usepackage{graphicx}
\usepackage{subcaption}

\usepackage{caption}
\usepackage{xcolor}

\usepackage{verbatim}

\usepackage[utf8]{inputenc}

\usepackage{setspace}
\usepackage{natbib}

\usepackage{lscape}

\usepackage{array}
\usepackage{arydshln}

\usepackage{abstract}

\usepackage{dcolumn}
\newcolumntype{,}{D{.}{.}{2}}

\usepackage[bottom]{footmisc}
\makeatother

\usepackage[skip=4pt, indent=0pt]{parskip}

\begin{document}
\setlength\parindent{0pt}
\thispagestyle{empty}
\renewcommand*{\thefootnote}{\fnsymbol{footnote}}

\section*{\centering{Digital technologies and performance incentives:\\
Evidence from businesses in the Swiss economy\footnote{This research has been funded by the Swiss National Science Foundation as part of the National Research Program NRP77 “Digital Transformation” under grant
No. 187462. The authors are solely responsible for the analysis and the interpretation thereof.}}}

\bigskip

\subsubsection*{Johannes Lehmann\footnote{Correspondence:\\
Johannes Lehmann\\
j.lehmann@unibas.ch}\footnote{University of Basel, Faculty of Business and Economics, Basel, Switzerland.} and \centering{Michael Beckmann$^\dagger$\footnote{Institute for Employment Research (IAB), Nuremberg, Germany; Institute of Labor Economics (IZA), Bonn, Germany. michael.beckmann@unibas.ch.}}}

\bigskip

\begin{center} \textit{Forthcoming in the Swiss Journal of Economics and Statistics}
\end{center}

\bigskip
\bigskip

\begin{abstract}
\bigskip

Using novel survey data from Swiss firms, this paper empirically examines the relationship between the use of digital technologies and the prevalence of performance incentives. We argue that digital technologies tend to reduce the cost of organizational monitoring through improved measurement of employee behavior and performance, as well as through employee substitution in conjunction with a reduced agency problem. While we expect the former mechanism to increase the prevalence of performance incentives, the latter is likely to decrease it. Our doubly robust ATE estimates show that companies using business software and certain key technologies of Industry 4.0 increasingly resort to performance incentives, suggesting that the improved measurement effect dominates the employee substitution effect. In addition, we find that companies emerging as technology-friendly use performance incentives more frequently than their technology-averse counterparts. Both findings hold for managerial and non-managerial employees. Our estimation results are robust to a variety of sensitivity checks and suggest that Swiss businesses leverage digital technologies to enhance control over production or service processes, allowing them to intensify their management of employees through performance incentives.

\bigskip
\bigskip

\textbf{Keywords:} Digital technologies, computer technologies, business software, key technologies of Industry 4.0, performance incentives, doubly robust ATE estimation
\end{abstract}

\newpage
\pagenumbering{arabic}
\setcounter{page}{1}


\renewcommand*{\thefootnote}{\arabic{footnote}}
\setcounter{footnote}{0}

\section{Introduction}
\label{section:Int}

Together with globalization and demographic change, digital transformation is one of the biggest challenges in current business landscapes.\footnote{Digital transformation can be defined as the use of digital technologies to adapt established patterns such as business processes, organizational structures and resources, or relationships with internal and external stakeholders \citep[][]{brynjolfssonComputationInformationTechnology2000, loebbeckeReflectionsSocietalBusiness2015, vialUnderstandingDigitalTransformation2019,  plekhanovDigitalTransformationReview2023}.} Economic research has long been studying the effects of technological change and continues to do so in the age of digitalization. For example, empirical research in labor economics has dealt with technology-induced employment effects \citep[e.g.,][]{autorWhyAreThere2015, autorAutomationLaborShareDisplacing2018, acemogluRobotsJobsEvidence2020, dixonRobotRevolutionManagerial2021}, skill-biased or routine-biased technological change \citep[e.g.,][]{bresnahanInformationTechnologyWorkplace2002b, goosExplainingJobPolarization2014, michaelsHasICTPolarized2014}, or issues of job and skill polarization \citep[e.g.,][]{autorSkillContentRecent2003, autorPolarizationLaborMarket2006, goosExplainingJobPolarization2014, autorWhyAreThere2015, autorAutomationLaborShareDisplacing2018}. Furthermore, empirical studies in organizational economics and management strategy have investigated the impact of new technologies on firm performance \citep[e.g.,][]{bresnahanInformationTechnologyWorkplace2002b, aralThreeWayComplementaritiesPerformance2012a, tambeExtrovertedFirmHow2012a, brynjolfssonArtificialIntelligenceModern2018}, on corporate and sourcing strategies \citep[e.g.,][]{abramovskyOutsourcingOffshoringBusiness2006a, acemogluVerticalIntegrationTechnology2010, aralInformationTechnologyRepeated2018}, on organizational and job design \citep[e.g.,][]{bresnahanInformationTechnologyWorkplace2002b, acemogluTechnologyInformationDecentralization2007, bloomDistinctEffectsInformation2014b, gertenControllingWorkingCrowds2019, gertenInformationCommunicationTechnology2022}, as well as on performance pay \citep[e.g.,][]{dixonRobotRevolutionManagerial2021, zwysenPerformancePayEurope2021, bayo-morionesComputerUsePay2022}. 

In our paper, we build on the the latter two literature strands, thereby referring to the empirical debate on the three-legged stool approach of organizational architecture developed in \citet[chapter 11]{brickleyManagerialEconomicsOrganizational2021}. This concept introduces the organizational architecture of a company as a system consisting of three interdependent subsystems (stool legs): the system of decision-rights assignment, the performance measurement system, and the reward system. While the organizational and job design literature looks at the technology effects on the systems of decision-rights assignment and performance measurement, and the performance pay studies consider the technology effects on the reward system, we combine the performance measurement and the reward systems of organizational architecture to investigate the effects of digital technologies on a system of performance incentives. Specifically, we examine whether the usage of computer technologies, business software solutions, and key technologies of Industry 4.0 in organizations is related to the prevalence of performance incentives, defined as the percentage of managerial and non-managerial employees who are subject to performance targets, performance evaluations, and pay-for-performance plans.

In principle, the effect of digital technologies on performance incentives can be positive or negative, where in both cases, technology usage contributes to reduce the cost of organizational monitoring, i.e., the cost of monitoring employee behavior or performance. In the first case, digital technologies are assumed to improve or simplify the measurement of employee behavior or performance \cite[]{lemieuxPerformancePayWage2009, dixonRobotRevolutionManagerial2021, zwysenPerformancePayEurope2021}. We refer to this scenario as the \textit{improved measurement effect}. In the second case, digital technologies are assumed to automate executing and monitoring tasks and are thus likely to replace both the supervising workers and the workers to be monitored, thereby additionally mitigating the conventional agency problem. This scenario is referred to as the \textit{employee substitution effect} \citep[]{ dixonRobotRevolutionManagerial2021}. We expect digital technologies to be positively related to the prevalence of performance incentives if the improved measurement effect dominates the employee substitution effect. In the reverse case, we expect a negative relationship.

To empirically test the digital technologies-performance incentives relationship, we use a new cross-sectional survey data set: the \textit{Swiss Employer Survey} (SES). The SES has been collected as a primary data set by ourselves using a sample from the Swiss Federal Statistical Office that is representative for Swiss firms with ten or more employees. To increase the number of observations, we supplemented the SES with our own sample drawn by means of web scraping. The final data set consists of 446 observations. Methodologically, we rely on a selection-on-observables approach using the doubly robust ATE estimator that combines inverse probability weighting with regression models for the potential outcome equations.

Our empirical results show a positive relationship between the use of digital technologies and the prevalence of performance incentives in Swiss companies. Most importantly, except for management information systems, all other forms of business software (i.e., groupware, enterprise resource planning, document management systems and customer relationship management) turn out to be positively associated with the prevalence of performance incentives. The same applies to artificial intelligence / big data solutions, cloud computing / storage and virtual boardrooms from the Industry 4.0 key technologies category. However, cyber-physical systems, the Internet of Things and robotics are found to be unrelated to the prevalence of performance incentives. These results suggest that most business software technologies and some key technologies of Industry 4.0 contribute to reduce the cost of organizational monitoring, where the improved measurement effect dominates the employee substitution effect. Furthermore, we find that companies identified as technology-friendly in their industry and size class are more likely to use performance incentives than their technology-averse counterparts, which is consistent with our previous findings. Finally, our estimation results indicate that managerial and non-managerial employees do not significantly differ in terms of exposure to technology-related performance incentives, suggesting that digital technologies reduce the cost of organizational monitoring in a similar manner across hierarchical levels. 

Our estimation results are robust to a variety of sensitivity checks, including the use of alternative measures for our composite variable of technological affinity, different weighting and trimming strategies, and a purely data-driven approach of covariate selection. Nevertheless, due to the cross-sectional nature of our data set, it is unlikely that we can interpret our doubly robust ATE estimates in terms of causal inference. However, we are confident that they are more informative and meaningful than conventional OLS estimates. In any case, our estimation results are in line with \cite{dixonRobotRevolutionManagerial2021}, \cite{zwysenPerformancePayEurope2021}, and \cite{bayo-morionesComputerUsePay2022}, who all focus on the effects of technology on performance pay and find positive associations with their technology variables.  

Our contribution to the aforementioned empirical literature on the relationship between technological innovation and the organizational architecture of the firm is as follows. First, our SES data set includes variables on the most advanced state-of-the-art technologies in digital transformation. In contrast to earlier studies, we thus have precise information on the incidence of a broad spectrum of cutting-edge digital technologies in Swiss companies. Second, the richness and novelty of our technology variables allows us to distinguish between technology-friendly and technology-averse firms. This gives us additional insight into the decisions made by companies that are at the forefront of managing digital transformation, including the challenges they face.\footnote{As far as we know, the only other paper that sheds light on the decisions and actions of technologically leading firms is \cite{acemogluTechnologyInformationDecentralization2007}. However, this paper is not about setting performance incentives but about the allocation of decision-making authority in firms.} Third, the SES data allow us to distinguish between managerial and non-managerial employees in the construction of the performance incentive variables. This enables us to identify potential heterogeneous, hierarchy-level-specific technology effects. This opportunity is of great interest in light of the fact that the work of (low-skilled) non-managerial workers is likely to be easier to monitor and evaluate than the work of (higher-skilled) managerial workers \cite[]{dixonRobotRevolutionManagerial2021}.\footnote{We otherwise observe separate technology effects for managerial and non-managerial workers only in the studies of \cite{gertenControllingWorkingCrowds2019, gertenInformationCommunicationTechnology2022}, who focus on mobile ICT (rather than a wide range of digital technologies) and use decentralized autonomy and centralized monitoring as outcome variables (rather than a system of performance incentives).} Finally, a fourth contribution of our paper is that we analyze the topic of digital transformation in the context of businesses in the Swiss economy. This is of particular interest and importance as, according to the Global Innovation Index 2022, Switzerland is the most innovative country in the world \citep[][p. 19]{duttaGlobalInnovationIndex2022}.

The remainder of this paper is structured as follows. Section \ref{section:theory} is devoted to the theoretical framework of our research question. Section \ref{section:DVIS} introduces the data and core variables and provides some descriptive statistics. In section \ref{section:methodology}, we describe our empirical methodology. In section \ref{section:balance_results}, we first present a graphical test of the common support assumption and some covariate balance diagnostics to get information on the quality of our model specification. We then discuss our estimation results, including the robustness checks. Finally, section \ref{section:Con} concludes.

\section{Theoretical framework}
\label{section:theory}
\begin{figure}[ht!]
\begin{center}
\caption{}
\label{fig:Monitoring}
\includegraphics[width=10cm, height=10.4cm]{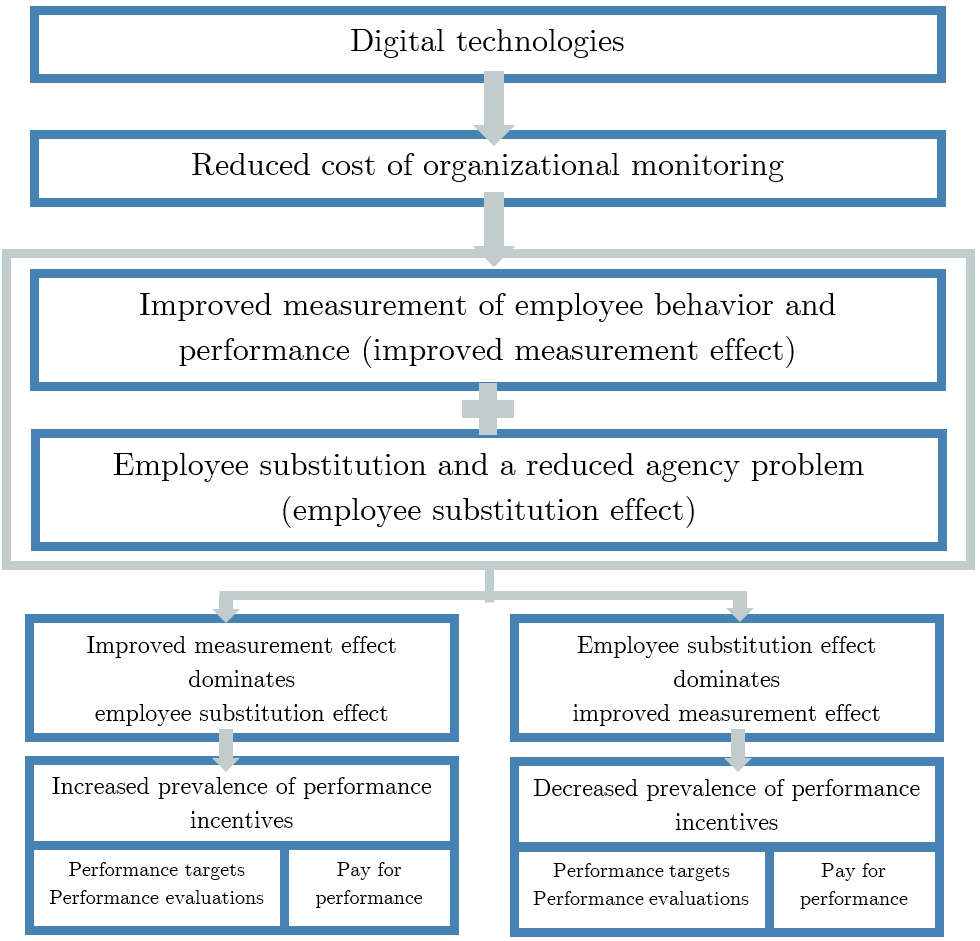}
\captionsetup{justification=centering,margin=1.5cm}
\end{center}
\end{figure}

In our study, we distinguish between three types of digital technologies: 
\begin{itemize}
    \item computer technologies, i.e., stationary and non-stationary ICT equipment (e.g., PCs, laptops, tablets, smartphones),
    \item business software, i.e., groupware (e.g., MS Teams, Zoom, Slack), enterprise resource planning (ERP), document management systems (DMS), customer relationship management (CRM), management information systems (MIS),
    \item key technologies of Industry 4.0, i.e., cyber-physical systems (CPS), Internet of Things (IoT), artificial intelligence (AI) / big data, cloud computing and storage, virtual boardrooms, robotics and automated transport or production systems, additive manufacturing / 3D print, virtual and augmented reality, blockchain.
\end{itemize}
Our theoretical considerations on the relation between the use of digital technologies and the prevalence of performance incentives are graphically illustrated in figure \ref{fig:Monitoring}.\footnote{A more detailed theoretical discussion can be found in section A1 in the online appendix.} 

The core argument is based on the idea that digital technologies help companies to reduce their cost of organizational monitoring. One reason for this is improved and simplified measurement of employee behavior and performance \cite[]{lemieuxPerformancePayWage2009, dixonRobotRevolutionManagerial2021, zwysenPerformancePayEurope2021}, which will make it more attractive for companies to set performance targets, carry out performance appraisals, and design pay-for-performance plans. For example, computer technologies and business software provides management with data that make performance measurement more accurate than before \cite[]{aralWhichCameFirst2006, hittInvestmentEnterpriseResource2002, aralThreeWayComplementaritiesPerformance2012a, collazosDescriptiveTheoryAwareness2019, koriatKnowledgeSharingAnalytics2019, bayo-morionesComputerUsePay2022, aladeDesignImplementationWebbased2023}, so managers can improve the measurement of employee input and output (computer technologies), learn about different levels of employee effectiveness across departments and hierarchies (ERP), track individual contributions of employees within project work (DMS), or obtain information about employee-customer interactions (CRM) as well as the dates and duration of online calls made by their employees (groupware). Likewise, Industry 4.0 technologies can contribute to improve and simplify the measurement of employee behavior and performance. For example, the division of labor between production workers and robots makes it easier to observe the performance attributable to production workers \cite[]{ dixonRobotRevolutionManagerial2021}. Furthermore, the combined use of AI and big data in personnel recruitment improves the measurement of skills, personality traits or performance of applicants, thus increasing the decision quality in recruitment processes \citep[][]{tambeArtificialIntelligenceHuman2019, giermindlDarkSidesPeople2022}. Also, cloud computing and storage provides a cost-effective way to store, access and manage data, including information about the behavior and performance of employees at work \cite[]{brauUtilizingPeopleAnalytics2023, choHumanResourcesAnalytics2023}. This is particularly true for software-as-a-service solutions that allow companies to track the working time of their employees and see in detail how employees spend their working day \cite[]{beckmannEntwicklungArbeitZeiten2018, nymanReformingWorkPatterns2023}. Similar to groupware, virtual boardrooms facilitate online meetings while collecting data about employees' work, such as time spent in meetings or meeting frequency \citep[e.g.,][]{gelbardSentimentAnalysisOrganizational2018, giermindlDarkSidesPeople2022}. Finally, increased data collection can be achieved through smart sensors and other IoT devices \citep[e.g.,][]{leeRecentAdvancesTrends2013, gaurStrengtheningPeopleAnalytics2019}.

A second reason for assuming that digital technologies help companies to reduce their cost of organizational monitoring can be attributed to the observation that digital technologies sometimes lead to the automation of production or service processes and an associated substitution of labor \cite[]{ dixonRobotRevolutionManagerial2021}. This is because automation technologies such as robots and AI have several comparative advantages over human labor, including higher productivity and quality (e.g., less variation in production and service processes, no interruptions at work, no breaks, no fatigue), lower costs (e.g., no strikes, no union membership, no wage increases, no paid vacation), and reduced agency problems. The latter result from a reduced need for supervision (fewer workers need to be supervised after automation, implying that fewer supervisors are needed) and from the fact that automation technologies do not engage in shirking behavior. Besides robots and AI, other Industry 4.0 technologies are also assumed to be associated with employee substitution, such as blockchain \citep[][]{manskiBuildingBlockchainWorld2017}, cyber-physical systems \citep[][]{waschullWorkDesignFuture2020} and additive manufacturing \citep[][]{adepoju3DPrintingAdditive2022, feliceEmploymentImplicationsAdditive2022}. However, it is important to note that employee substitution is likely to be limited to the subset of automation technologies, while we expect that all digital technologies considered have the potential to improve the measurement of employee behavior and performance.

Overall, therefore, the reduction in the cost of organizational monitoring is achieved in two ways: through improved measurement of employee behavior and performance as well as through employee substitution in conjunction with a reduced agency problem. If we refer to the first argument as the \textit{improved measurement effect} and the second as the \textit{employee substitution effect}, then our first hypothesis to be tested can be formulated as: 

\textit{Hypothesis} 1: \textit{The use of individual digital technologies is interrelated with the prevalence of performance incentives via an improved measurement effect and an employee substitution effect.} 
\begin{itemize}
\item[(a)] \textit{If the improved measurement effect dominates the employee substitution effect, digital technologies will be positively related to the use of performance incentives.}
\item[(b)] \textit{If the employee substitution effect dominates the improved measurement effect, digital technologies will be negatively related to the use of performance incentives.}
\end{itemize}

In addition to analyzing the association between the use of individual digital technologies and the prevalence of performance incentives, we examine whether companies that are in different stages of digital transformation also differ in terms of the prevalence of performance incentives. We thus no longer distinguish between companies in terms of their use of individual digital technologies, but rather in terms of the extent to which they use digital technologies. For this purpose, we look at two groups of companies: the technology-friendly and the technology-averse companies. We define companies as technology-friendly if they have implemented more digital technologies than the median of comparable companies. Consequently, technology-averse companies have implemented the same number or fewer digital technologies than the median of comparable companies. Our second hypothesis can then be written as: 

\textit{Hypothesis} 2: \textit{Technology-friendly and technology-averse companies differ with regard to the prevalence of performance incentives.} 
\begin{itemize}
\item[(a)] \textit{If the improved measurement effect dominates the employee substitution effect, tech\-nol\-ogy-friendly companies are more inclined to make use of performance incentives than their technology-averse counterparts.}
\item[(b)] \textit{If the employee substitution effect dominates the improved measurement effect, tech\-nol\-ogy-friendly companies are less inclined to make use of performance incentives than their technology-averse counterparts.}
\end{itemize}

\section{Data, variables, and descriptive statistics}
\label{section:DVIS}
\subsection{Data}
\label{section:Data}

Our empirical study is based on primary data collected from establishments\footnote{An establishment is understood as a physically separate organizational unit that operates independently or is part of a larger organization. In the further course of the paper, we will use the terms establishment, business, firm, and company equivalently.} in the Swiss economy providing information for the years 2020 or 2022, the \textit{Swiss Employer Survey} (SES). The addresses of the companies were provided to us by the Swiss Federal Statistical Office in the form of a sample that is representative of Swiss companies with ten or more employees. In total, we were able to contact 10,000 companies in this way and asked for their support of our research project.\footnote{To establish contact, we contacted the companies a maximum of two times by letter, with the second contact being intended as a reminder. The sample does not cover establishments of the public administration, farming, and mining sectors.} To ensure sufficient coverage of larger establishments with 250 or more employees, care was taken that this group was disproportionately represented in the drawing of the final sample. 

The SES covers a wide range of business topics, such as information on workforce structures, corporate cultures and strategies, as well as organizational architecture including decision-rights assignment, performance measurement and remuneration policies. In addition, the SES contains questions on the companies' financial situation as well as their business environment including their market situation and regulation issues. Other topics refer to staff recruiting, working time regimes and further training. The focus of the SES, however, is on where Swiss companies stand in terms of digital transformation. 

The response rate to our employer survey was very modest and is only about six percent (not cleaned of firms with very incomplete information). The original target was a 'typical' response rate of around 20\%. The timing of the survey coincided with the height of the Covid-19 pandemic, when Swiss businesses and companies were forced to continue operating under the most difficult conditions and thus certainly had other concerns than participating in our survey. After excluding establishments that provided very incomplete information, we end up with a sample size of 297 observations. To increase the number of observations, we supplemented this sample with a data set drawn via web scraping. In this way, we were able to increase the number of observations by about 150 companies. Thus, our final data set consists of 446 firms.

The web scraped sample was constructed as follows. First, a list of employer and industry associations was manually compiled. In a second step, the contact details of the member companies listed on the association's website were extracted either manually (for smaller associations) or automatically using a Python script (for larger associations). We contacted the generated sample by e-mail. To keep the survey population comparable to the baseline sample, we excluded establishments with less than 10 employees and establishments that already participated in the baseline survey.

To check whether the answers obtained from the two data sources are comparable, we depict the mean values of the variables of interest and the included control variables separately for these two subgroups in table A9 in the online appendix. Overall, the differences are very modest and lie in the range of a few percentage points. The average company from the web scraping sample is slightly less technology-friendly and relies less on the use of incentive systems. This is probably due to the fact that the average company in the web scraping sample is somewhat smaller than in our SES sample. Overall, the differences observed are not alarming and show that the businesses in the two subgroups have relatively similar characteristics.\footnote{Re-estimating the baseline specifications just with the sample of establishments provided by the Swiss Federal Statistical Office produces very similar results.}
Another problem could be the relatively low response rate and the fact that our survey took place during the Covid-19 pandemic. Whether this poses a problem for the validity of our estimation results can be assessed by comparing the core descriptive statistics of our survey with corresponding statistics from comparable data sets from Switzerland or abroad. We perform this comparison in section \ref{section:Desc}.

\subsection{Digital technologies variables}
\label{section:Tech_variables}

Our SES data contains a rich set of variables providing information on a series of business decisions in the context of digital transformation. A first set of technology variables refers to the usage of a number of contemporary digital technologies in Swiss businesses. In total, the SES includes 16 types of digital technologies, and companies were asked whether or not these technologies are being used. An overview of all 16 binary technology variables and the specific survey questions can be found in tables A11 and A12 in the online appendix. These dummies $DT_{1}, DT_{2}, ..., DT_{16}$ are considered as treatment variables to test \textit{Hypothesis} 1. 

To test \textit{Hypothesis} 2, we construct a composite variable $DTint$ providing information on the intensity of digital technologies usage. In a sense, we follow the widespread practice of proxying the level of digitalization in a firm by the available information on the use of specific technologies \citep[e.g.,][]{aralWhichCameFirst2006, aralThreeWayComplementaritiesPerformance2012a, gertenControllingWorkingCrowds2019, dixonRobotRevolutionManagerial2021, bayo-morionesComputerUsePay2022}. However, while the measurement of technological intensity in the empirical literature often relies on only one technology, we extend this approach by using all 16 binary technology indicators to construct an index of technological intensity, thus ensuring a comprehensive view of the state of digital transformation in Swiss companies.

For the construction of this index variable, we follow empirical studies, such as \cite{bresnahanInformationTechnologyWorkplace2002b}, \cite{bloomAreFamilyfriendlyWorkplace2011}, \citet[][]{gertenControllingWorkingCrowds2019, gertenInformationCommunicationTechnology2022}, and \cite{ beckmannEmpowermentTaskCommitment2022}, and apply a double-standardization approach in which each technology dummy variable is standardized before the sum of the standardized technology variables is standardized. The resulting technology intensity variable $DTint$ can thus be written as 
\begin{equation*}
DTint = STD\{STD(DT_{1}) + STD(DT_{2}) + ... + STD(DT_{16})\} \, .
\end{equation*}
By construction, $DTint$ has zero mean and unit variance. We utilize $DTint$ to construct a binary treatment variable that separates technology-friendly firms from their technology-averse counterparts, thereby applying the following two-step procedure. First, we divide the sample into three firm size classes (small, medium, large) and two industries (manufacturing, service sector) and subsequently assign each establishment to one of the resulting six cells. The distinction by industry and firm size ensures that establishments are compared to similar firms. In a second step, we define firms located above the median of the $DTint$ distribution in each cell as technology-friendly and firms located at or below the median of the $DTint$ distribution in each cell as technology-averse.

This two-step procedure provides us with a binary treatment variable indicating technology-friendly firms, $TFint$, which is defined as 
\begin{equation*}\label{TFint}
TFint = 
\begin{cases}
1 & \text{if } DTint > DTint^{0.5} \\
0 & \text{if } DTint \leq DTint^{0.5} \, ,
\end{cases}
\end{equation*}
where $DTint^{0.5}$ represents the median of the $DTint$ distribution in a specific industry-firm size cell. For methodological reasons that will be explained later in section \ref{section:diagnostics}, we will use ten out of the 16 digital technology dummies $DT_{1}$ through $DT_{16}$ and the dummy variable $TFint$ indicating technology-friendly firms as treatment variables in our baseline treatment effects models presented and discussed in section \ref{section:results}.

\subsection{Performance incentives variables}
\label{section:Incentive_variables}

Our SES data set contains three measures of performance incentives: the shares of employees covered by performance target agreements ($Target$), performance evaluations ($Eval$), and performance pay plans ($Pay$). This information is available separately for managerial ($m$) and non-managerial ($nm$) employees.\footnote{The specific survey questions used to construct the dependent variables are displayed in table A12 in the online appendix. For the performance pay variable, we have information on the prevalence at three different hierarchical levels. To approximate the prevalence at the managerial level, we calculated the mean value for the top and the middle/lower management positions.} After applying the double-standardization approach, our main dependent variable for performance incentives $Inc^{j}$ can be written as
\begin{equation*}
Inc^{j} = STD\{STD(Target^{j}) + STD(Eval^{j}) + STD(Pay^{j})\} \, ,
\end{equation*}
where $j \in \{m,nm\}$. By construction, $Inc^{j}$ has zero mean and unit variance. We will use $Inc^{j}$ as well as its standardized components as dependent variables in our treatment effects analyses.

\subsection{Control variables}
\label{section:CV}

In the context of a selection-on-observables approach, the choice of control variables is of particular importance. In our analysis, we combine theoretical and statistical considerations to select an appropriate set of covariates. From the statistical point of view, we follow \citet{imbensNonparametricEstimationAverage2004}, \citet{austinIntroductionPropensityScore2011}, \citet{liUsingPropensityScore2013}, \citet{ austinMovingBestPractice2015}, and \citet{naritaCausalInferenceObservational2023}, who recommend to control for all (true) confounding covariates, i.e., variables that jointly determine the dependent and the treatment variable, and for potential confounders (prognostically important covariates), i.e., covariates that determine the outcome variable but not necessarily the treatment variable. Including true confounders helps us to reduce selection bias, while potential confounders may increase the accuracy of the estimated treatment effect without harming identification. However, we do not make use of control variables related to the treatment variable but not to the outcome variable, sometimes referred to as instruments, because these covariates will increase the variance of the estimated treatment effect even if identification is not affected.

Our research question can be incorporated into a theoretical framework developed in \citet[chapter 11]{brickleyManagerialEconomicsOrganizational2021}. In this framework, the organizational architecture of a firm is determined by the firm's business environment in terms of the technological, market and regulatory situation, and the firm's strategies. Now that our treatment variable covers the technology domain of a firm's business environment and our dependent variable reflects two stool legs of organizational architecture, i.e., performance measurement and reward, our control variables must capture the remaining dimensions of this theoretical framework, i.e., the firms' market and regulatory environment, the strategies employed, and the system of decision-rights assignment. 

To capture the market component of business environment, we control for the number of competitors as a measure of competitive pressure and for the seven Swiss greater regions (e.g., Northwestern Switzerland, Zurich, Central Switzerland, Ticino) to reflect different local demand conditions. The regulatory component of business environment is measured by the legal form of a company (private vs. capital company), the existence of an organizational unit for employee representation (similar to a works council or firm-level union), and a dummy variable indicating whether or not a company is legally independent or part of a larger organization. We map the strategy domain by an industry dummy separating the manufacturing from the service sector\footnote{We recognize that the distinction between the manufacturing and the service sector reflects very general categories. While the estimated performance incentives effects tend to be somewhat smaller when we include information on sixteen industries according to the 2008 NOGA classification instead of the sector dummy, the level of statistical significance does not change. Moreover, the level of achieved balances between the treatment and control groups decreases when this information is added. The combination of these two facts motivates the decision to include only the dummy variable separating the manufacturing from the service sector in the baseline regression models. The lasso estimates presented in section A2.5 in the online appendix contain finer information about the NOGA classification of an establishment.} and by two dummy variables indicating a company's make-or-buy strategies (internal and external expansion strategies, business unit sales and outsourcing decisions) in the past five years. We control for the decision-rights assignment dimension of organizational architecture by adding a double-standardized centralization-decentralization variable, which is composed of nine items (work planning, definition of tasks, quality control of work, replenishment of raw and auxiliary materials, pace of work, sequence of work, contact with customers, investment in machinery or equipment, as well as granting of compensation, bonuses and promotion of employees) providing information on whether decisions are made by non-managerial workers, first-line and middle managers, or top management. Finally, we control for company size by three dummy variables (small, medium-sized, large), for the skill structure (share of high-, medium-, low skilled employees) within companies, and for the fact that our survey took place at two different points in time.\footnote{Table A9 in the online appendix presents the descriptive statistics of all included control variables.} 

The construction of the theoretical framework described in \citet[chapter 11]{brickleyManagerialEconomicsOrganizational2021} ensures the selection of true and potential confounders while excluding the instruments. We check the robustness of this approach of selecting control variables in section A2.5 in the online appendix, where we apply a pure data-driven approach to select the covariates for our estimation models using double-selection lasso linear regression. Based on our complete data set, we let lasso select the true confounders for one specification and then relax the restrictions on selecting the covariates for another specification, for which we let lasso additionally select the instruments and potential confounders.

\subsection{Descriptive statistics}
\label{section:Desc}

The descriptive statistics presented in this section provide first insights with respect to the incidence of digital technologies and performance incentives in Swiss companies. To ensure that the descriptive statistics calculated from the SES are representative of the Swiss business sector, we apply sampling weights. The weighting process takes into account the seven Swiss greater regions, sixteen industries according to the NOGA classification 2008, and three company size classes.

\subsubsection{Treatment variables}
\label{section:Desc_treat}

\begin{figure}[ht!]
\begin{center}
\caption{Incidence of digital technologies in Swiss companies}
\label{fig:tech}
\includegraphics[width=0.9\linewidth]{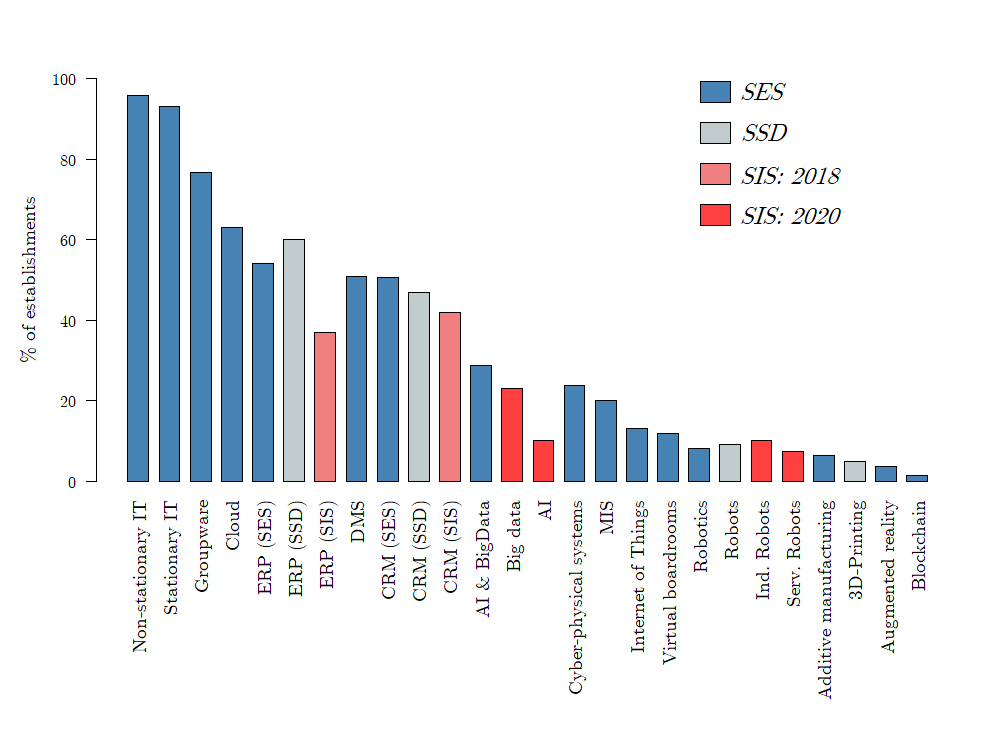}
\label{Anz_tech}
\end{center}
\caption*{\emph{Source}: Swiss Employer Survey (SES); own calculations.\\
\emph{Notes}: Number of observations: 446. Calculations include sample weights.}
\end{figure}

Referring to our SES survey, figure \ref{fig:tech} depicts the prevalence of sixteen digital technologies across Swiss establishments. Unsurprisingly, computer technologies, i.e., non-stationary and stationary IT, are implemented in almost all Swiss establishments (96\% and 93\%, respectively). The various forms of business software are now also quite widespread in Swiss companies, ranging between about 50\% (ERP, CRM, DMS) and 75\% (groupware), with the outlier for MIS at 20\%. The relatively high prevalence of groupware can probably be explained by the occurrence of the Covid-19 pandemic, because many companies purchased software such as MS Teams or Zoom during the lock-down periods in order to maintain communication. 

Among the key technologies of Industry 4.0 examined, cloud computing / storage is the most widespread, with around 60\% of user companies, followed by AI / big data solutions and CPS. Around one in four companies uses these technologies. IoT and virtual boardrooms are only used in around one in ten companies. The prevalence of all other Industry 4.0 technologies, i.e., robotics and automated transport or production systems, additive manufacturing, augmented reality, and blockchain, is below 10\% . It is somewhat surprising that the key technologies of Industry 4.0 are used very little in some cases. Terms such as IoT, 3D print, or blockchain are omnipresent today, but apparently mainly in public discussions and less in practical implementation. 

To check whether any concerns about the relatively low response rate to the SES or the timing of its launch during the Covid-19 pandemic are justified, we compare the penetration rates of digital technologies in our SES with the corresponding rates observed in two other Swiss firm-level surveys: the Swiss Survey on Digitalization (SSD) from autumn 2016 and the Swiss Innovation Survey (SIS), which is conducted every two years \citep[e.g.,][]{beckAnalyseDigitalisierungSchweizer2020a, speschaInnovationSchweizerPrivatwirtschaft2020, speschaInnovationUndDigitalisierung2023a}. Due to the fact that the survey populations differ and the definitions of the technology variables are sometimes not identical, a direct comparison of the results needs nuanced interpretation. For example, the SIS covers establishments with more than five employees, the SES with more than ten employees, and the SSD with more than 20 employees. This is important as larger establishments are generally more likely to implement digital technologies.

The following penetration rates result for the digital technologies surveyed in the three data sets: 
\begin{itemize}
    \item ERP: SES 54\%, SSD 60\%, SIS 37\%
    \item Robotics: SES 8\%, SSD 9\% (industry robots), SIS 10\% (industry robots), 8\% (service robots)
    \item CRM: SES 51\%, SSD 47\%, SIS 42\%
    \item Additive manufacturing: SES 6\%, SSD 5\%
    \item AI / big data: SES 29\%, SIS 10\% (AI), 23\% (big data)
\end{itemize}

Overall, it can be stated that the penetration rates of digital technologies in Swiss companies are quite close to each other across the domestic surveys, whereby smaller differences can essentially be explained by different target populations and survey periods. The only major difference is with SIS, where the penetration rate for ERP is only 37\%. This difference to our SES can be explained primarily by the fact that the SIS includes very small companies with fewer than ten employees, and the SES does not. The difference in the target populations also explains why ERP has the highest penetration rate in the SSD compared to the SES (and the SIS) despite an earlier survey date. This data comparison provides suggestive evidence that our SES data is not limited due to the low response rate or the survey launch during the Covid-19 pandemic.\footnote{Another comparable data set is the linked personnel panel (LPP), where the employer survey collects information about German establishments with more than 50 employees. With regard to the incidence of digital technologies, the LPP reports the following percentages for 2020: big-data analytics: 23\%, AI: 7\%, additive manufacturing: 13\%, cyber-physical systems: 21\%, robotics: 12\%, virtual or augmented reality: 4\% (the percentages can be found in the counts for the LPP variables; see https:\/\/fdz.iab.de\/int\_bd\_pd\/linked-personnel-panel-lpp-version-1221-v2\/).} Nonetheless, in section \ref{section:weighting}, we address potential concerns of non-response bias due to a low response rate by using sampling weights not only for the descriptive statistics but also for the regression analyses.

\subsubsection{Outcome variables}
\label{section:Desc_out}

\begin{figure}[ht]
\centering
\caption{Prevalence of performance incentives in Swiss companies}
\label{fig:incentives}
\includegraphics[width=1\linewidth]{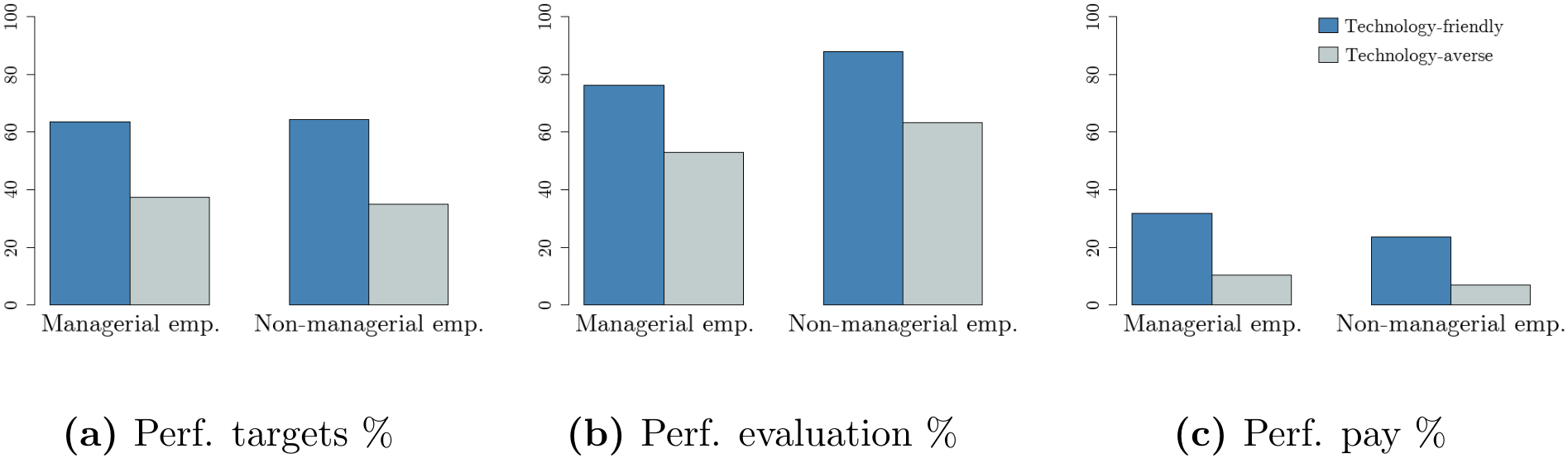}
\smallskip
\caption*{\emph{Source}: Swiss Employer Survey (SES); own calculations.\\
\emph{Notes}: Number of observations: 446. Calculations include sample weights.}
\end{figure}

Turning to our dependent variables, i.e., the three types of performance incentives, we observe some divergence in their prevalence. Specifically, 85\% of the surveyed establishments apply performance evaluations, 65\% utilize performance targets, and 37\% employ pay-for-performance schemes. Figure \ref{fig:incentives} illustrates this divergence at a more granular level showing the proportions of employees subject to performance targets, performance appraisals and performance pay plans in an average establishment. The figure further distinguishes between managerial and non-managerial employees as well as between technology-friendly and technology-averse companies, as defined by our treatment variable $TFint$.

Performance evaluations, the most common of the three incentive schemes, are applied for 66\% of the managerial employees and 77\% of the non-managerial employees in the average establishment. In addition, performance targets apply to an identical proportion of 51\% at both hierarchy levels. Finally, pay-for-performance plans are the least widely implemented management practice, used for 22\% of the managerial employees and 16\% of non-managerial employees.\footnote{The German LPP employer survey collects similar information for the average establishment (with more than 50 employees) in 2020. Here, it is reported that 41\% of managerial and 15\% of non-managerial employees are subject to performance targets. Furthermore, 58\% of managerial and 51\% of non-managerial employees are subject to performance evaluations, and 19\% of non-managerial employees receive pay-for-performance (the proportion of managerial employees subject to performance pay is not reported; the percentages can be approximated from the counts for the LPP variables; see https:\/\/fdz.iab.de\/int\_bd\_pd\/linked-personnel-panel-lpp-version-1221-v2\/). These numbers suggest, that the application of performance incentives is more widespread in Switzerland than in Germany, especially with regard to performance evaluations.} In summary, this means that managerial and non-managerial employees are quite equally affected by performance targets. On the other hand, performance appraisals apply slightly more to non-managerial employees, while performance pay is used primarily for managerial employees. Such differences across employees located at different hierarchical levels lead us to the idea that the use of digital technologies may have heterogeneous and hierarchy-specific effects on the prevalence of performance incentives in organizations. 

With regard to the prevalence of performance incentives in technology-friendly and tech\-nol\-ogy-averse companies, it can be observed that all forms of performance incentives are significantly more widespread in technology-friendly companies than in technology-averse companies. This holds true for both managerial and non-managerial employees. This finding can be viewed as a first indication of a positive relationship between the use of digital technologies and the prevalence of performance incentives in Swiss companies.

\begin{figure}[ht]
\centering
\caption{Intensity of performance incentives grouped by number of technologies}
\label{fig:incentives_tech}
\includegraphics[width=0.8\linewidth]{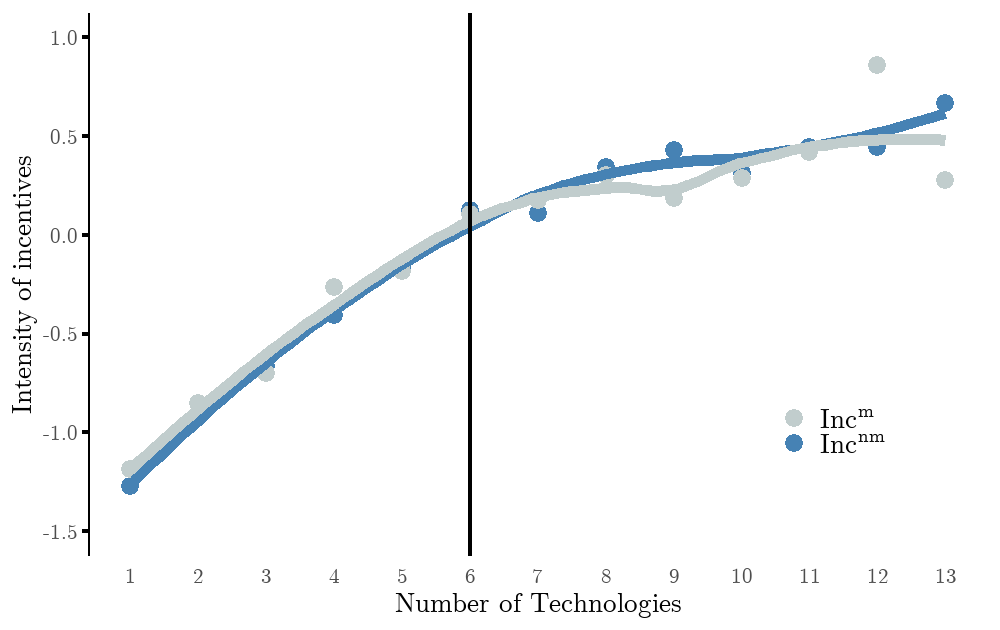}
\smallskip
\caption*{\emph{Source}: Swiss Employer Survey (SES); own calculations.\\
\emph{Notes}: Number of observations: 444. 
We excluded two observations as outliers because only one establishment each makes use of 14 and 15 technologies, respectively.}
\end{figure}

Figure \ref{fig:incentives_tech} graphically illustrates further information on the relationship between the intensity of technology usage and the prevalence of performance incentives. The x-axis shows the number of digital technologies used in the average company. On the y-axis, we see the values of the variables $Inc^m$ and $Inc^{nm}$. The figure displays the averages of $Inc^m$ and $Inc^{nm}$ within the group of establishments that make use of a certain number of technologies. These data points are shown as dots in the figure. The figure also shows two lines representing the trends in the prevalence of performance incentives for different numbers of technologies applied. The black vertical line represents the median of the applied technologies in the sample, which is six digital technologies.

Figure \ref{fig:incentives_tech} paints a very similar picture for both hierarchy levels, i.e., managerial and non-managerial employees. In both cases, the figure shows a monotonously increasing course of the performance incentives curves, implying that the more digital technologies a company uses, the more intensive is the use of performance incentives. It appears that both curves are steeper on the left hand side of the median than on the right hand side. At first, the curves are virtually linear, but to the right of the median the curves flatten out somewhat. This underlines that the choice of the median as the threshold value is not only theoretically natural, but is also indicated by the empirical picture. However, the most important insight from this graphical analysis is that the descriptive finding supports the validity of Hypothesis 2 a), according to which the improved measurement effect dominates the employee substitution effect, so that technology-friendly companies are more inclined to make use of performance incentives than their technology-averse counterparts. 

It is important to note that these descriptive results provide only an initial impression of the relationship between digital technologies and performance incentives in Swiss companies. More substantial and informative insights can be obtained through treatment effect regression analyses.\footnote{Table A9 in the online appendix presents further descriptive statistics of the outcome and treatment variables.}

\section{Empirical methodology}
\label{section:methodology}

\subsection{Identification strategy}
\label{section:IDS}

The starting point of our empirical analysis is the potential outcome model
\begin{equation}
PI_{1,i}^{j} = X_{i} \beta_{1} + U_{1,i} \quad \text{ if } DT_{i} = 1
\label{eq:PO1}
\end{equation}
\begin{equation}
PI_{0,i}^{j} = X_{i} \beta_{0} + U_{0,i} \quad \text{ if } DT_{i} = 0 \, ,
\label{eq:PO0}
\end{equation}
where $PI_{1}^{j}$ and $PI_{0}^{j}$ represent our dependent variables on performance incentives introduced in section \ref{section:Incentive_variables}, i.e., $Target^{j}$, $Eval^{j}$, $Pay^{j}$ in their standardized forms, and the composite variable $Inc^{j}$. The index $i$ denotes the respective firm, and the superscript $j$ characterizes managerial employees $(j=m)$ or non-managerial employees $(j=nm)$. $DT$ is a binary treatment variable indicating a firm's utilization of one of the specific digital technologies introduced in section \ref{section:Tech_variables} as well as the dummy variable $TFint$ separating technology-friendly from technology-averse firms. $X$ is the matrix of control variables discussed in section \ref{section:CV}, while the vectors $\beta_{1}$ and $\beta_{0}$ represent the parameters to be estimated. Finally, $U_{1}$ and $U_{0}$ denote stochastic error terms with zero mean and finite variance.

We are interested in estimating the average treatment effect (ATE) of digital technology utilization on the prevalence of performance incentives in Swiss companies, i.e., $\widehat{PI}_{1,i} - \widehat{PI}_{0,i}$. For this purpose, we apply a doubly robust estimation strategy for the ATE developed in \cite{robinsEstimationRegressionCoefficients1994}. The doubly robust ATE estimator combines inverse probability weighting (IPW) with regression models for the potential outcome equations (\ref{eq:PO1}) and (\ref{eq:PO0}). 

ATE estimation using the doubly robust estimator is a three-step approach. The first step is a probit maximum likelihood estimation of the parameters of the treatment probability model
\begin{equation}
    p(X_{i}) = Pr(DT = 1 \, | \, X = X_{i}) = DT_{i} = \Phi(X_{i} \beta) \, ,
    \label{eq:treat}
\end{equation}
where $\Phi$ is the cumulative distribution function of the standard normal distribution.
From these estimates, we compute the propensity score $\hat{p}(X_{i})$ and the IPWs \cite[]{austinIntroductionPropensityScore2011, austinMovingBestPractice2015}.

In a second step, we estimate the potential outcome models (\ref{eq:PO1}) and (\ref{eq:PO0}) 
and predict the treatment-specific outcomes $\widehat{PI}_{1,i}^{j} = x'_{i} \hat{\beta}_{1}$ and $\widehat{PI}_{0,i}^{j} = x'_{i} \hat{\beta}_{0}$ for the entire sample to impute the unobserved counterfactual \citep[][p. 1292]{cameronMicroeconometricsUsingStata2022}. In a third step, we finally calculate the weighted means of the predicted potential outcomes. 

The doubly robust ATE is then computed by the difference of these weighted averages \citetext{\citealp[e.g.,][]{funkDoublyRobustEstimation2011}; \citealp[]{abdiaPropensityScoresBased2017}}.

This three-step procedure provides consistent estimates that can be interpreted in terms of causal inference if the stable unit treatment value assumption (SUTVA), the conditional independence assumption (CIA), and the common support assumption (CSA) are satisfied \citep[e.g.,][]{imbensNonparametricEstimationAverage2004,  austinIntroductionPropensityScore2011, liUsingPropensityScore2013, abdiaPropensityScoresBased2017, naritaCausalInferenceObservational2023}.

For our empirical analysis, CIA is the most critical assumption. This is because at the present stage of data availability, we can only draw on cross-sectional data. However, in order to convincingly estimate causal ATEs within the scope of a selection-on-observables approach such as doubly robust ATE estimation, the availability of a comprehensive panel data set including pre-treatment control variables and post-treatment outcome variables would be required.

Although it appears unlikely that the ATEs resulting from our doubly robust estimator can be interpreted in terms of causal inference, the doubly robust property makes this estimator superior to parametric OLS or other semiparametric estimators such as IPW. The doubly robust property means that the estimator provides consistent ATEs even if either the potential outcomes framework (\ref{eq:PO1}) and (\ref{eq:PO0}) or the IPW model (\ref{eq:treat}) is incorrectly specified \citetext{\citealp[]{funkDoublyRobustEstimation2011}; \citealp[]{abdiaPropensityScoresBased2017}}. In contrast, OLS and IPW require correct specification of the assumed functional form for the outcome model (OLS) or the treatment probability model (IPW).

\subsection{Weighting and trimming procedures}
\label{section:weighting}

Our empirical methodology incorporates the use of sampling weights to ensure that our ATE estimates are representative for the population of Swiss businesses with at least ten employees. In addition, we use sampling weights to mitigate potential non-response bias that may result from low response rates, as in our case. The weight adjustment process calibrates the surveyed sample to the Swiss business population across the seven Swiss greater regions, sixteen industries classified under the NOGA 2008 system, and three company size categories.

Since the use of sampling weights in the context of regression analysis is discussed quite controversially in the literature \cite[]{solonWhatAreWe2015, bollenAreSurveyWeights2016, busemeyerAuthoritarianValuesWelfare2022}, we check whether our ATE estimates are sensitive to sample weighting by reporting both weighted and unweighted estimates. The comparison of weighted and unweighted ATE estimates serves as a useful test against model misspecification \cite[]{brickUnitNonresponseWeighting2013, solonWhatAreWe2015}. Section \ref{section:results} shows the results of our baseline ATE estimations using sample weights, while the corresponding ATE estimates without sample weights are relegated to section A2.4 in the online appendix and serve as a robustness check. 

The CSA ensures that the estimated IPWs do not grow too large. Nevertheless, treated companies with a propensity score close to zero and non-treated companies with a propensity score close to one can receive very large IPWs. Extreme weights can also be generated by the sample weighting procedure when a firm belongs to an under-sampled part of the firm population. If, as in our case, both IPW and sample weighting are combined, the presence of extreme weights can be exacerbated because both weights are multiplied with each other. The existence of extreme weights impairs the precision of the doubly robust ATE \cite[]{austinMovingBestPractice2015, naritaCausalInferenceObservational2023}. In such cases, trimming extreme weights can effectively reduce the sampling variance of ATE estimates. 

Since trimming weights always entails a trade-off\footnote{Trimming can reduce the sampling variance and enhance the precision of the ATE. However, this comes at the cost of an increased mean square error and generally leads to a deterioration in the achieved covariate balance \citep[e.g.,][]{coleConstructingInverseProbability2008, potterMethodsIssuesTrimming2015}.}, there are various viable options to set the cut-off percentiles for trimming. Table \ref{tab:Weights} presents the viable options that are most frequently recommended in the literature \citep[e.g.,][]{sturmerTreatmentEffectsPresence2010, leeWeightTrimmingPropensity2011, thoemmesPrimerInverseProbability2016, hashimotoTheoryPracticePropensity2022} as well as the summary statistics for the weights calculated with $TFint$ as dependent variable in the treatment model (\ref{eq:treat}).

\begin{table}[!ht] \centering \footnotesize

\caption{\textbf{Summary statistics for the weights generated for $TFint$}}
\label{tab:Weights}

\begin{tabularx}{\textwidth}{l*{3}{>{\centering\arraybackslash}X}}
\toprule
\toprule
                                      & Min.     & Mean    & Max.  \\
                                      &(1)       &(2)      &(3) \\\midrule
No trimming                           & 0.170    & 2.024   & 22.106 \\
1st and 99th percentile               & 0.209    & 1.982   & 10.413 \\
2.5 and 97.5 percentile               & 0.246    & 1.910   & 7.565  \\
5th and 95th percentile               & 0.303    & 1.817   & 5.406  \\ 
\bottomrule
\bottomrule
\end{tabularx}
\begin{tablenotes}
\small
\item
\emph{Source}: Swiss Employer Survey (SES); own calculations. \\
\end{tablenotes}

\end{table}

To ensure that the chosen weights are neither too large nor too small, we opt for the 5th and 95th percentiles as the cut-off points for our baseline models. We investigate whether this choice affects our estimation results in a robustness check described in section A2.4 in the online appendix, where we use alternative trimming cut-off percentiles.

\section{Empirical results}
\label{section:balance_results}

\subsection{Common support and balance diagnostics}
\label{section:diagnostics}

\begin{figure}[ht]
\centering
\caption{\textbf{Propensity score densities}}
\label{fig:CommonSupport}
\includegraphics[width=0.6\linewidth]{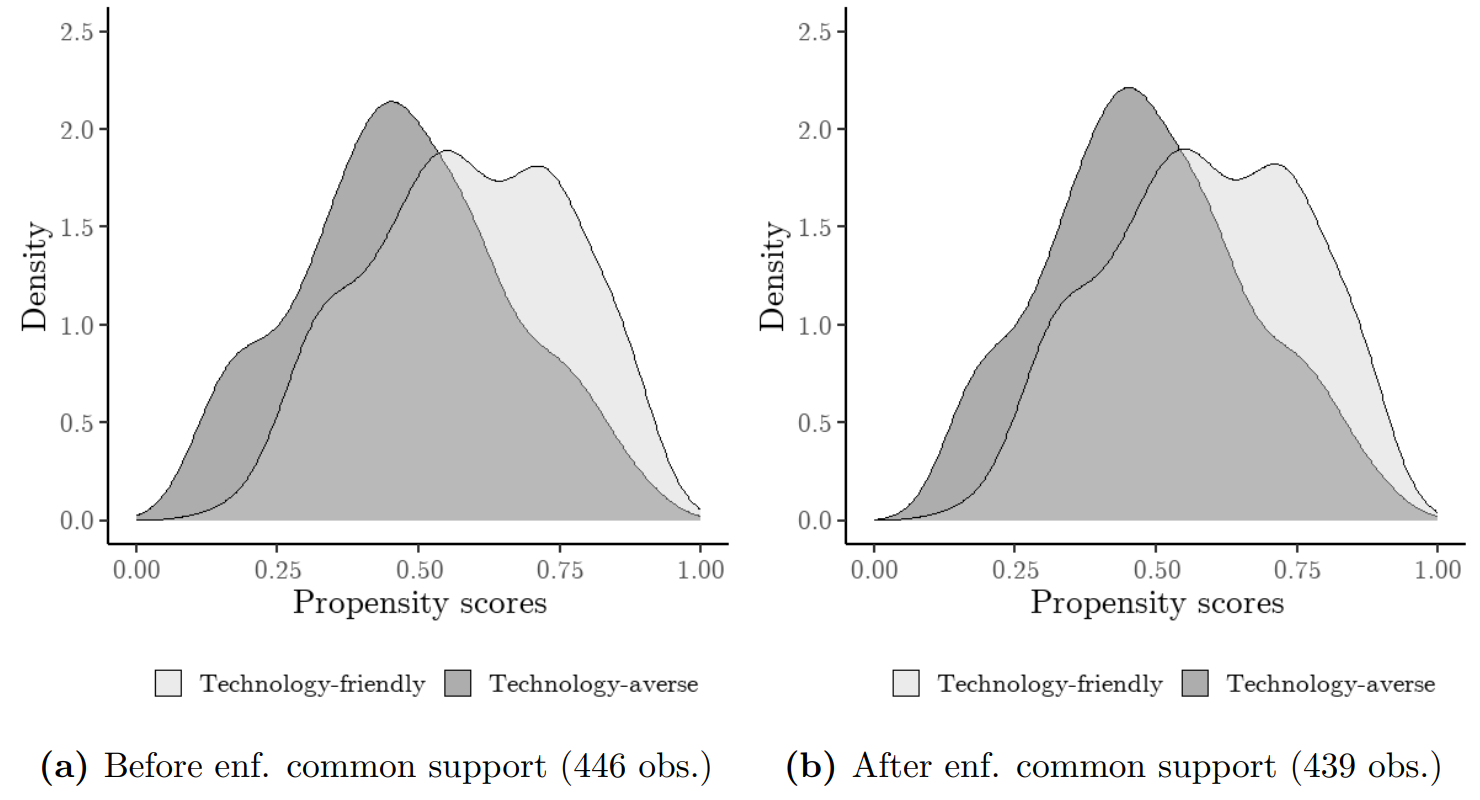}
\smallskip
\caption*{\emph{Source}: Swiss Employer Survey (SES); own calculations.}
\end{figure}

This section presents supplementary evidence on the adequacy of the propensity score model (\ref{eq:treat}). To achieve this, we assess the degree of common support and check whether IPW successfully reduces covariate imbalances between the treatment and control groups. Figure \ref{fig:CommonSupport} displays the estimated densities of the predicted probabilities that a technology-friendly company is indeed technology-friendly and that a technology-averse company is actually technology-friendly both before and after enforcing common support. The two density functions overlap over almost the entire propensity score interval, even before enforcing common support. As a result, enforcing common support only leads to the exclusion of seven observations, and the subsequent probability density functions are almost identical to their initial counterparts. We thus conclude that CSA is satisfied.

To quantify the balance of the observed covariates between treated and non-treated companies, we calculate the absolute standardized mean differences (ASMDs).\footnote{The ASMDs are probably the most widely used method of balance diagnostics \citep[e.g.,][]{austinIntroductionPropensityScore2011, liUsingPropensityScore2013, austinMovingBestPractice2015, abdiaPropensityScoresBased2017, naritaCausalInferenceObservational2023}.} If the ASMD detects systematic differences even after conditioning on the propensity score, this may indicate that the propensity score model is not correctly specified \cite[]{austinIntroductionPropensityScore2011}.\footnote{There is no clear consensus in the literature on the exact tolerable threshold that indicates a prevailing imbalance. Often, a threshold of $ASMD < 0.1$ is required, while other studies allow for $ASMD < 0.2$ \citep[e.g.,][]{austinIntroductionPropensityScore2011, austinMovingBestPractice2015, abdiaPropensityScoresBased2017}.} For $TFint$, the balance diagnostics suggest that the covariates are very similar across the treatment and control groups after weighting. The highest ASMD is 0.079, which is below the conservative threshold of 0.1.\footnote{Table A7 in the online appendix depicts the covariate balance statistics for all treatment variables of interest.} Regarding the single technologies, persistently high ASMDs exist when technology usage is either extremely prevalent (stationary and non-stationary ICT) or very rare (robotics, additive manufacturing, augmented reality, blockchain). For this reason, we cannot test \textit{Hypothesis} 1 for computer technologies and the aforementioned Industry 4.0 technologies. However, since $Robotics$ plays an important role in the derivation of \textit{Hypothesis} 1 b), we will present the corresponding OLS estimates to provide suggestive evidence.

For the further technologies, the application of IPW substantially improves the achieved balance. The highest ASMD is below 0.1 in five cases ($DMS$, $CRM$, $AI/BigData$, $Cloud$, and $CPS$), and below 0.2 in three further cases ($Groupware$, $ERP$, $MIS$). As IPW significantly improved the balance of covariates for $IoT$ and $Virtual\ boardrooms$, we keep these technologies in the analysis, even though their highest ASMDs are 0.25 and 0.22, respectively. We will interpret their ATE estimates with particular caution.

\subsection{Baseline regression results}
\label{section:results}

Table \ref{tab:MainRegression} reports the ATE estimates resulting from the doubly robust estimator. Columns (1)-(8) refer to different dependent variables capturing the different components of performance incentives. While columns (1), (3), (5), and (7) refer to the relationship between digital technology usage and the prevalence of performance incentives for managerial employees ($m$), columns (2), (4), (6), and (8) show the corresponding ATE estimates for non-managerial employees ($nm$). Columns (1) and (2) display the estimated ATEs for our composite outcome variable $Inc$. The remaining columns present the corresponding ATE estimates for each component of performance incentives, i.e., performance evaluations ($Eval$) in columns (3) and (4), performance targets ($Target$) in columns (5) and (6), and performance pay ($Pay$) in columns (7) and (8). Each row refers to a different treatment variable. The upper panel depicts the outcomes for the various business software applications, while the subsequent panel illustrates the results pertaining to the key technologies of Industry 4.0. The panel at the bottom of the table presents the findings for $TFint$, i.e., the composite technology variable separating technology-friendly from technology-averse firms. For each ATE estimate presented, we specified and ran a separate regression model. Therefore, table \ref{tab:MainRegression} reports the estimation results of 96 distinct regressions. 

Our two test hypotheses allow for both positive and negative relationships between the use of digital technologies and the prevalence of performance incentives, depending on whether the improved measurement effect dominates the employee substitution effect or vice versa. If anything, we expect the employee substitution effect to dominate over the improved measurement effect in the key technologies of Industry 4.0 rather than in business software applications.

For the business software solutions category, the doubly robust ATE estimates reveal a strong positive association with the composite performance incentive variables $Inc^m$ and $Inc^{nm}$. Only the results for $MIS$ turn out to be statistically insignificant. When looking at the results for the individual components of performance incentives, it is noticeable that the most statistically significant effects on the $Inc$ variables come from the ATEs for the performance pay variables $Pay^m$ and $Pay^{nm}$. Finally, there is no significant difference in the results for managerial and non-managerial employees, suggesting that the technology effect on performance incentives is not substantially different across hierarchical levels. Overall, the results are consistent with \textit{Hypothesis} 1 a), suggesting that the use of business software is generally associated with improved measurement of employee behavior and performance.

In the category of the key technologies of Industry 4.0, the connection with the prevalence of performance incentives is less clear than for business software solutions. Here, we find significant positive ATEs only for the variables $AI/BigData$, $Cloud$, and $VirtBoard$, while the corresponding ATEs for $CPS$, $IoT$ and $Robotics$ turn out to be statistically insignificant. The econometric results therefore support \textit{Hypothesis} 1 a) only for the combined use of AI and big data, cloud computing and storage, as well as for virtual boardrooms (whereby the results for the latter technology could suffer from the not entirely satisfactory balance statistics). In addition, we do not find any evidence for the dominance of the employee substitution effect for any of the key technologies of Industry 4.0 considered, as none of the estimated ATEs turns out to be significantly  negative. Finally, as with business software solutions, we find no evidence of different ATE effects across hierarchical levels.

The estimates for the composite technology treatment variable $TFint$ are also positive and highly statistically significant. The transition from a technology-averse company to a technology-friendly company is associated with an increase in the prevalence of performance incentives of 0.625 standard deviations among managerial employees and 0.707 standard deviations among non-managerial employees. The corresponding ATE estimates of $TFint$ on the individual performance incentive practices show somewhat smaller effects, ranging between 0.432 and 0.582, but are still highly statistically significant.\footnote{A more descriptive interpretation of the technology treatment effects is obtained if, instead of the standardized dependent variables $Eval$, $Target$, and $Pay$, we use their non-standardized counterparts as outcome variables measured as percentages. To give just one example, an ATE for $TFint$ on the standardized performance evaluation variable $Eval^{m}$ of 0.452 standard deviations corresponds to a percentage change of 19.5\%, meaning that a transition from a technology-averse to a technology-friendly company is associated with a 19.5\% increase in the proportion of managers who regularly receive performance appraisals, which is quite considerable.} Once again, no substantial differences in the estimates across hierarchy levels can be identified, suggesting that digital technologies reduce the cost of organizational monitoring in a similar manner across hierarchical levels. These empirical results therefore provide statistical evidence for the validity of \textit{Hypothesis} 2 a).

\begin{landscape}
\begin{table} \centering \footnotesize

\caption{\textbf{Doubly robust ATE estimates of digital technology usage on the diffusion of performance incentives}}
\label{tab:MainRegression}

\begin{tabularx}{\linewidth}{ll*{2}{>{\centering\arraybackslash}X}|*{2}{>{\centering\arraybackslash}X}|*{2}{>{\centering\arraybackslash}X}|*{2}{>{\centering\arraybackslash}X}}
\toprule
\toprule
&& $Inc^{m}$ & $Inc^{nm}$ & $Eval^{m}$ & $Eval^{nm}$ & $Target^{m}$ & $Target^{nm}$ & $Pay^{m}$ & $Pay^{nm}$ \\
&&(1)       &(2)       &(3)       &(4)       &(5)       &(6)       &(7)    &(8)\\ \midrule
&Groupware & 0.441*** & 0.470*** & 0.186 & 0.281* & 0.416*** & 0.368*** & 0.393*** & 0.368***  \\ 
&~ & (0.128) & (0.130) & (0.159) & (0.151) & (0.134) & (0.133) & (0.102) & (0.113)  \\ 
&ERP & 0.374*** & 0.432*** & 0.298** & 0.385*** & 0.146 & 0.152 & 0.402*** & 0.398***  \\ 
&~ & (0.127) & (0.127) & (0.129) & (0.111) & (0.123) & (0.126) & (0.118) & (0.129)  \\ 
&DMS & 0.500*** & 0.533*** & 0.349*** & 0.375*** & 0.434*** & 0.386*** & 0.345*** & 0.392***  \\ 
&~ & (0.117) & (0.120) & (0.119) & (0.108) & (0.121) & (0.124) & (0.114) & (0.122)  \\ 
&CRM & 0.588*** & 0.621*** & 0.369*** & 0.401*** & 0.514*** & 0.516*** & 0.445*** & 0.426***  \\ 
&~ & (0.112) & (0.112) & (0.121) & (0.102) & (0.113) & (0.116) & (0.113) & (0.121)  \\
&MIS & 0.028 & 0.047 & 0.076 & 0.145 & 0.060 & 0.028 & -0.073 & -0.071  \\ 
\multirow{-10}{*}{\rotatebox{90}{Business software}}  &        ~ & (0.136) & (0.129) & (0.143) & (0.111) & (0.127) & (0.130) & (0.138) & (0.143)  \\ \midrule
&AI/BigData & 0.392*** & 0.513*** & 0.228* & 0.424*** & 0.354*** & 0.392*** & 0.304** & 0.294**  \\ 
&~ & (0.121) & (0.119) & (0.127) & (0.102) & (0.124) & (0.123) & (0.125) & (0.132)  \\ 
&CPS & 0.080 & 0.111 & 0.163 & 0.338*** & -0.012 & -0.008 & 0.030 & -0.091  \\ 
&~ & (0.129) & (0.127) & (0.137) & (0.113) & (0.129) & (0.125) & (0.128) & (0.128)  \\ 
&IoT & 0.098 & 0.236 & 0.100 & 0.101 & -0.012 & 0.206 & 0.132 & 0.204  \\ 
&~ & (0.141) & (0.152) & (0.155) & (0.144) & (0.145) & (0.148) & (0.154) & (0.163)  \\ 
&Cloud & 0.252** & 0.294** & 0.191 & 0.245** & 0.105 & 0.170 & 0.274** & 0.220*  \\ 
&~ & (0.115) & (0.117) & (0.122) & (0.111) & (0.115) & (0.119) & (0.109) & (0.113)  \\ 
&VirtBoard & 0.231* & 0.275** & 0.477*** & 0.465*** & 0.076 & 0.252* & -0.030 &  -0.121  \\ 
&~ & (0.122) & (0.120) & (0.112) & (0.089) & (0.156) & (0.149) & (0.137) & (0.146)\\

&Robotics$^\dagger$ & -0.084 & 0.180 & -0.205 & 0.049 & -0.060 & 0.080 & 0.076 & 0.261 \\ 
\multirow{-12}{*}{\rotatebox{90}{\parbox{3cm}{\centering Key technologies of Industry 4.0}}} & ~ & (0.194) & (0.200) & (0.221) & (0.217) & (0.194) & (0.185) & (0.173) & (0.714)
\\  \midrule
&$TFint$  & 0.625*** & 0.707*** & 0.452*** & 0.516*** & 0.482*** & 0.581*** & 0.477*** & 0.432***  \\ 
&~ & (0.112)  & (0.111) & (0.121) & (0.106) & (0.114) & (0.114) & (0.108) & (0.119)  \\ 

\bottomrule
\bottomrule
\end{tabularx}
\begin{tablenotes}
\small
\item
\emph{Source}: Swiss Employer Survey (SES); own calculations. \\
\emph{Notes}: *, **, and *** represent statistical significance at the 10\%, 5\%, and 1\% level, respectively. Each entry in the table refers to a distinct estimation. The values in parentheses represent robust standard errors. The calculations include sample weights and IPW. Table A11 lists the abbreviations for the digital technologies. The number of observations and balance statistics are denoted in table A7. Table A8 depicts the full regression results for the regressions of $Inc^{m}$ and $Inc^{nm}$ on $TFint$ and the covariates. All these tables are part of the online appendix. $^\dagger$The regression coefficients of \emph{Robotics} stem from an OLS regression without IPW.
\end{tablenotes}

\end{table}
\end{landscape}

Overall, therefore, the ATEs resulting from our doubly robust estimation approach provide strong evidence for the validity of \textit{Hypothesis} 1 a) and \textit{Hypothesis} 2 a), according to which digital technologies reduce the cost of organizational monitoring through improved measurement of worker activities and employee substitution, where the former effect dominates the latter. This even applies to $AI/BigData$, which we basically viewed not only as a technology that contributes to better performance measurement, but also as automation technology with a certain potential to replace employees. In none of our estimation models do we find evidence of the dominance of the employee substitution effect over the improved measurement effect. Instead, the improved measurement effect very often dominates the employee substitution effect, suggesting that in Swiss businesses using digital technologies, the focus is on improving control over the production or service process, but not on replacing workers. Moreover, we find no indication of differential technology effects on the prevalence of performance incentives across hierarchical levels. This result does not support the validity of the assumption of \citet{dixonRobotRevolutionManagerial2021}, according to which work at lower hierarchical levels is easier to monitor than work at higher hierarchical levels. Nevertheless, our ATE estimates are consistent with the findings obtained in \citet{dixonRobotRevolutionManagerial2021}, \citet{zwysenPerformancePayEurope2021}, and \citet{bayo-morionesComputerUsePay2022}, who all report positive associations between technology usage and pay for performance.

\subsection{Sensitivity analysis}
\label{section:sen}

To ensure the robustness of our results, we perform five sensitivity analyses. While section A2 in the online appendix provides a comprehensive discussion, this section summarizes the approaches and results. All robustness checks focus on the treatment variable $TFint$.

First, we re-specify $TFint$ by proxying digital transformation with alternative variables (section A2.1). The estimated coefficients remain significant in almost all cases, except for the $Eval^{nm}$ regression, in which the doubly robust ATE is marginally statistically insignificant ($p$ = 0.111). Second, we apply different thresholds in the dichotomization of $DTint$ (section A2.2). Positive ATEs emerge for technological leaders (top 25th percentile) and negative ATEs for technological laggards (bottom 25th percentile). Both results indicate a positive relationship between the use of digital technologies and the prevalence of performance incentives. Third, we re-estimate our baseline regression using IPW instead of the doubly robust estimator (section A2.3). Since the estimated ATEs are very similar to those obtained for the baseline model, there is no indication of model misspecification. Fourth, we report the results with adjusted trimming and weighting procedures (section A2.4).  The estimated ATEs remain statistically significant when we omit sampling weights or choose different cutoffs in the trimming procedure, indicating that the specific choices regarding weighting or trimming do not drive the observed effects. Finally, we apply a purely data-driven approach to select the control variables (section A2.5). While the obtained ATEs turn out to be somewhat smaller than their counterparts in the baseline specifications, they remain highly statistically significant.

Overall, all five sensitivity checks support the validity of our baseline results, which indicate a positive relationship between the application of digital technologies and the prevalence of performance incentives.

\section{Conclusion}
\label{section:Con}

In this paper, we empirically examine the relationship between the utilization of digital technologies and the prevalence of performance incentives in Swiss companies. For this purpose, we analyze novel observational data from Swiss establishments at the cross-sectional level: the \textit{Swiss Employer Survey} (SES). 

From the theoretical framework we conclude that digital technologies reduce the cost of organizational monitoring through two mechanisms: improved measurement of employee behavior and performance (improved measurement effect) and employee substitution in conjunction with a reduced agency problem (employee substitution effect). A dominant improved measurement effect predicts a positive association between the use of digital technologies and the prevalence of performance incentives, while a dominant employee substitution effect predicts a negative relationship. We test these opposing predictions by making use of a doubly robust ATE estimation approach that combines inverse probability weighting (IPW) with regression models for the potential outcome equations. 

Our estimation results provide evidence for the prevailing dominance of the improved measurement effect. None of the empirical results points in the direction of a dominating employee substitution effect. Specifically, we find that almost all business software solutions (i.e., groupware, enterprise resource planning (ERP), document management systems (DMS), and customer relationship management (CRM)), as well as some key technologies of Industry 4.0 (i.e., AI / big data solutions, cloud computing and storage, and virtual boardrooms) turn out to be positively related with the prevalence of performance incentives. Remarkably, we do not find any statistically significant ATEs for cyber-physical systems (CPS) or the Internet of Things (IoT). A first explanation could be that both the improved measurement effect and the employee substitution effect occur here, which offset each other, so that neither effect dominates the other (this may also explain the statistically insignificant effect for robotics). However, it is also conceivable that these two technologies are used very heterogeneously in companies. Ultimately, they are seen as generic terms for the organization of production in the age of digitalization. Furthermore, we find that technology-friendly companies are more likely to use performance incentives than their technology-averse counterparts. Finally, our estimation results do not reveal significant differences for  managerial and non-managerial employees, suggesting that digital technologies reduce the cost of organizational monitoring in a similar manner across hierarchical levels. Our baseline estimates are robust to a variety of sensitivity checks, including the use of alternative measures for technological affinity, different strategies to binarize our treatment variable, lasso covariate selection, and adjusted weighting and trimming procedures.

Our doubly robust ATE estimates are consistent with the results obtained in \cite{dixonRobotRevolutionManagerial2021}, \cite{zwysenPerformancePayEurope2021}, and \cite{bayo-morionesComputerUsePay2022}, who find positive associations between new technologies and the use of performance pay. Our findings are also consistent with the results of studies on the performance impact of management systems consisting of complementary management practices \citep[e.g.,][]{aralThreeWayComplementaritiesPerformance2012a, wuDataAnalyticsSupports2019, wuDataAnalyticsInnovation2020}. For example, the authors of the first study mentioned identify complementarities between human resource analytics, human capital management (a part of the ERP business software), and pay for performance, which is consistent with the positive relationship between digital technologies (including ERP and AI / big data analytics) and performance incentives (including pay for performance) obtained in our study. Finally, our doubly robust ATE estimates are in line with the results obtained in studies investigating the impact of ICT on job design 
\citep[e.g.,][]{ gertenControllingWorkingCrowds2019, gertenInformationCommunicationTechnology2022} and pay for performance plans \citep[][chapter 6]{gertenThreeEssaysOrganizations2022}. These studies find positive technology effects on centralized monitoring through performance appraisals on the one hand, and on the usage of collective performance pay plans on the other hand. Among others, important differences to our work exist in the technology variables (mobile ICT vs. a large set of contemporary digital technologies, including mobile ICT) and in the methodological estimation approaches.

Overall, our empirical results suggest that Swiss companies implement digital technologies to improve control over production and service processes. If digital technologies improve the performance measurement within production and service processes, this also includes the provision of additional or more accurate data to measure employee behavior and performance. This in turn encourages companies to intensify their usage of performance incentives.   

One limitation of our study is that our doubly robust ATE estimates probably cannot be interpreted in terms of causal inference because we cannot rule out the presence of unobserved confounding or reverse causation. This is because our SES data set is so far only available at the cross-sectional level. To convincingly estimate causal effects in a selection-on-observables setting, we would require information from both the treatment period and the pre- and post-treatment periods. On the other hand, with our estimation strategy we already make some efforts to reduce the endogeneity problem of our treatment variables. In this respect, we consider our estimation results to be more meaningful than conventional OLS estimates.  

\bibliography{References.bib}

\newpage

\section*{Declarations}
\subsection*{Availability of data and materials}
The data for the empirical analysis is available in an anonymous form upon reasonable request from Johannes Lehmann.

\subsection*{Competing interests}
The authors declare that they have no competing interests.

\subsection*{Funding}
We would like to thank the Swiss National Science Foundation who finances our research project on \textit{The Swiss labour market in the digital transformation (SWISSLAB)} as part of the National Research Program NRP 77 \textit{Digital Transformation}, grant number 187462, for the generous support. We are also grateful to the Swiss Federal Statistical Office for providing us with a representative sample of Swiss companies, which forms the basis for our survey and subsequent data analysis.

\subsection*{Authors' contributions}
Both authors contributed to the data collection, conceptual idea and writing of the paper. JL was responsible for the data processing, empirical estimations, and visualizations. Both authors have read and approved the final version of the manuscript.

\subsection*{Acknowledgements}
Not applicable

\newpage
\subsection*{Abbreviations}

\begin{table*}[ht!]
\begin{tabular}{ll} \toprule\toprule
AddMan      &	    Additive manufacturing processes\\
AI/BigData &	    Software or algorithms for IT-based process optimization\\
ASMD        &		Absolute standardized mean difference\\
ATE         &       Average treatment effect\\
AugReality  &	    Virtual/augmented reality\\

Block	    &	    Blockchain\\

CIA	        &	    Conditional independence assumption\\
Cloud       &		Cloud storage/computing\\
CPS	        &	    Cyber-physical systems\\
CRM         &		Customer relationship management\\
CSA         &       Common support assumption\\

DMS	        &	    Document management systems\\

ERP	        &	    Enterprise resource planning\\
Eval        &       Performance evaluations\\

Groupware   &	    Groupware/Collaborative applications\\

IoT	        &	    Internet of Things\\
IPW         &       Inverse probability weighting\\

m           &       Mangerial employees\\
MIS	        &	    Management information systems\\

nm          &       Non-managerial employees\\
NOGA        &       Nomenclature générale des activitée economiques \\
&(General classification of Economic activities)\\
NonStat     &	    Non-stationary IT-equipment\\

Pay         &       Pay-for-performance\\
PDF	        &  	    Probability density function\\

Robotics    &  	    Robotics, automated transport or production systems\\

SD          &       Standard deviation\\
SES         &       Swiss Employer Survey\\
SIS         &       Swiss Innovation Survey\\
SSD         &       Swiss Survey on Digitalization\\
Stat        &		Stationary IT-equipment\\
SUTVA       &		Stable unit treatment value assumption\\
Sw.	        & 	    Switzerland\\

Target      &       Performance targets\\

VirtBoard   & 	   Virtual Boardrooms\\

\bottomrule\bottomrule
\end{tabular}
\end{table*}

\end{document}